\documentclass[usenatbib]{mn2e}
\usepackage{natbibmnfix,graphicx,times}
\usepackage{amsbsy}

\newcommand{\bq}{\begin{equation}}
\newcommand{\eq}{\end{equation}}
\newcommand{\bqa}{\begin{eqnarray}}
\newcommand{\eqa}{\end{eqnarray}}

\newcommand{\eV}{\mbox{ eV}}
\newcommand{\kel}{\mbox{ K}}
\newcommand{\mkel}{\mbox{ mK}}

\newcommand{\hunits}{\mbox{ km s$^{-1}$ Mpc$^{-1}$}}

\newcommand{\recunits}{\mbox{ cm$^{3}$ s$^{-1}$}}

\newcommand{\Junits}{\mbox{ cm$^{-2}$ s$^{-1}$ Hz$^{-1}$ sr$^{-1}$}}

\newcommand{\bxhi}{\bar{x}_{\rm HI}}
\newcommand{\xhi}{x_{\rm HI}}

\newcommand{\bxion}{\bar{x}_i}

\newcommand{\dtb}{\delta T_b}
\newcommand{\bdtb}{\bar{\delta T}_b}

\newcommand{\lya}{Ly$\alpha$ }

\newcommand{\bk}{{{\bf k}}}

\newcommand{\deriv}{{\rm d}}

\newcommand{\apj}{ApJ}
\newcommand{\apjl}{ApJ}

\newcommand{\aap}{A\&A}
\newcommand{\aj}{AJ}
\newcommand{\mnras}{MNRAS}

\newcommand{\prd}{PRD}
\newcommand{\pra}{PRA}
\newcommand{\prl}{PRL}

\title[Electron-hydrogen spin exchange rates]{Spin exchange rates in electron-hydrogen collisions}

\author[S.~R. Furlanetto \& M.~R. Furlanetto]{Steven R.  Furlanetto$^1$\thanks{Email:  steven.furlanetto@yale.edu} \& Michael R. Furlanetto $^2$\thanks{Email:  mfurlanetto@lanl.gov}\\
$^1$Yale Center for Astronomy and Astrophysics, Yale University, PO Box 208121, New Haven, CT 06520-8121\\
$^2$Physics Division, Los Alamos National Laboratory, MS H803, P.O. Box 1663, Los Alamos NM 87545}

\begin{document}

\maketitle

\begin{abstract}
The spin temperature of neutral hydrogen, which determines the 21 cm
optical depth and brightness temperature, is set by the competition
between radiative and collisional processes.  In the high-redshift
intergalactic medium, the dominant collisions are typically those
between hydrogen atoms.  However, collisions with electrons couple much more efficiently to
the spin state of hydrogen than do collisions with other hydrogen
atoms and thus become important once the ionized fraction exceeds
$\sim 1 \%$.  Here we compute the rate at which electron-hydrogen
collisions change the hydrogen spin.  Previous calculations included
only $S$-wave scattering and ignored resonances near the $n=2$
threshold.  We provide accurate results, including all partial wave
terms through the $F$-wave, for the de-excitation rate at temperatures $T_K \la 1.5
\times 10^4 \kel$; beyond that point, excitation to $n \ge 2$
hydrogen levels becomes significant.  Accurate electron-hydrogen collision rates at higher temperatures
are not necessary, because collisional excitation in this regime
inevitably produces \lya photons, which in turn dominate spin
exchange when $T_K \ga 6200 \kel$ even in the absence of radiative
sources. Our rates differ from previous calculations by several
percent over the temperature range of interest.  We also consider
some simple astrophysical examples where our spin de-excitation
rates are useful.
\end{abstract}

\begin{keywords}
atomic processes -- scattering -- diffuse radiation
\end{keywords}

\section{Introduction} \label{intro}

The 21 cm transition is potentially a powerful probe of the
pre-reionization intergalactic medium (IGM) because of the enormous
amount of neutral hydrogen in the Universe at that time
\citep{field58, scott90, madau97}.  It can teach us about
reionization, the formation of the first structures and the first
galaxies, and even the ``dark ages" before these objects formed
(\citealt{furl06-review}, and references therein).  It is therefore
crucial to understand the fundamental physics underlying the 21 cm
transition.  One critical aspect is the spin temperature, which is
determined by the competition between the scattering of cosmic microwave
background (CMB) photons, the scattering of \lya photons
\citep{wouthuysen52, field58}, and collisions.  When CMB scattering
dominates, the IGM remains invisible because the spin temperature
approaches that of the CMB (which is used as a backlight).

Before star formation commences, collisions are the only way to
break this degeneracy.  The total coupling rate is determined by
collisions both with other hydrogen atoms and with electrons.  At the low
residual electron fraction expected after cosmological recombination
\citep{seager99}, H--H collisions dominate.  Spin exchange in such
interactions has received a great deal of attention over the years
\citep{purcell56, smith66, allison69, zygelman05, hirata06}.  Such
collisions suffice to couple the spin temperature $T_S$ to the
kinetic temperature $T_K$ at high densities, when $z \ga 50$
\citep{loeb04}.  However, the spin exchange rate in these collisions
falls dramatically at $T_K \la 100 \kel$; coupled with the declining
density, such interactions become ineffective at lower redshifts.
H--e$^-$ collisions have received much less attention. However, at
fixed temperature, free electrons move at much higher velocities
than do hydrogen atoms, and are thus much more efficient at spin exchange
on a per-particle basis.  If the mean ionized fraction is $\bxion
\ga 0.01$, they can play a significant role in the coupling. This
situation can occur if, for example, X-rays from the first quasars
or star-forming galaxies create a warm, partially-ionized medium
\citep{oh01, venkatesan01, ricotti04_a, ricotti05, nusser05-xray,
kuhlen05-sim, kuhlen06-21cm}, or if the decay or annihilation of
exotic particles deposits a significant amount of energy into the
IGM \citep{chen04-decay, pierpaoli04, shchekinov06, furl06-dm}.

The spin de-excitation rate in H--e$^-$ collisions, which we will
denote $\kappa_{10}$, obviously depends on the detailed scattering
properties of these collisions.  Over the past several decades, this
interaction has received a great deal of attention as a model
problem for electron-atom interactions (e.g., \citealt{massey51,
schwartz61, temkin62, burke62a, poet78, bray02}).  At IGM
temperatures $T_K \la 10^4 \kel$, most electrons have energies below
the threshold to excite the $n=2$ level, and the dominant mechanism
of spin de-excitation is electron-electron spin exchange.  Although
this regime has been particularly well-studied, it has only been
solved in full nonrelativistic detail relatively recently.  The
latest spin exchange calculation \citep{smith66} used the scattering
phase shifts computed numerically by \citet{schwartz61}.  While
accurate, these included only $S$-wave scattering and neglected both
higher-order partial waves and H$^-$ resonances near the $n =2$
threshold (which were not resolved by the numerical algorithm). Both
of these issues have been solved over the intervening four decades
(see, e.g., \citealt{wang93, wang94} for a recent calculation).  Our
goal is to recalculate $\kappa_{10}$ in light of these newer phase
shifts.

Higher-order partial waves become significant at $T_K \ga 1000
\kel$; if X-ray heating in the early Universe is strong, then these
temperatures can be achieved easily \citep{furl06-glob}.  We
therefore examine the high-temperature limit in some detail.  Our
calculation breaks down once collisional electronic excitation (and
ionization) become important.  We will show that, above $T_K \sim
1.5 \times 10^4 \kel$, direct spin de-excitation by scattering above
the $n=2$ threshold cannot be ignored.  Although such interactions
are difficult to model -- because of the wide array of possible
transitions -- in practice this regime is relatively unimportant. In
the low-density IGM, collisional excitation to the $2p$ and higher
levels is followed rapidly by radiative de-excitation, producing a
\lya background.  We show that this background inevitably dominates
the spin temperature coupling at $T_K \ga 6200 \kel$ (assuming a
thermal distribution of electrons).  The crossover can occur at
smaller temperatures if a non-thermal population of fast electrons
exists (produced, for example, by X-rays; \citealt{chen06,
chuzhoy06-first}).

The remainder of this paper is organized as follows.  In \S
\ref{21cm}, we briefly review the 21 cm transition.  Our main
results are contained in \S \ref{ehcoll}, where we examine various
mechanisms for spin de-excitation, consider the high-temperature
limit, and calculate the spin de-excitation rates for H--e$^-$
collisions.  We present some simple astrophysical applications in \S
\ref{astro}, and we conclude in \S \ref{disc}.

In our numerical calculations, we assume a cosmology with
$\Omega_m=0.26$, $\Omega_\Lambda=0.74$, $\Omega_b=0.044$, $H=100 h
\hunits$ (with $h=0.74$), $n=0.95$, and $\sigma_8=0.8$, consistent
with the most recent measurements \citep{spergel06}, although we
have increased $\sigma_8$ from the best-fit \emph{WMAP} value in
order to improve agreement with weak-lensing data.

\section{The 21 cm Transition} \label{21cm}

We review the relevant characteristics of the 21 cm transition here;
we refer the interested reader to \citet{furl06-review} for a more
comprehensive discussion.  The 21 cm brightness temperature
(relative to the CMB) of a patch of the IGM is
\begin{eqnarray}
\dtb & = & 27 \, \xhi \, (1 + \delta) \, \left( \frac{\Omega_b h^2}{0.023} \right) \left( \frac{0.15}{\Omega_m h^2} \, \frac{1+z}{10} \right)^{1/2} \nonumber \\
& & \times \left( \frac{T_S - T_\gamma}{T_S} \right) \, \left[ \frac{H(z)/(1+z)}{\deriv v_\parallel/\deriv r_\parallel} \right] \mkel,
\label{eq:dtb}
\end{eqnarray}
where $\delta$ is the fractional overdensity, $\xhi$ is
the neutral fraction, $x_i = 1 - \xhi$ is the ionized fraction, $T_S$ is the
spin temperature, $T_\gamma$ is the CMB temperature, and $\deriv
v_\parallel/\deriv r_\parallel$ is the gradient of the proper
velocity along the line of sight.  The last factor accounts for
redshift-space distortions  \citep{bharadwaj04-vel,barkana05-vel}.

The spin temperature $T_S$ is determined by competition between
scattering of CMB photons, scattering of UV photons
\citep{wouthuysen52, field58}, and collisions \citep{purcell56}.  In
equilibrium,\footnote{This is an excellent approximation throughout
cosmic history.  The worst case occurs when interactions with CMB photons dominate;
the relevant timescale is then $t_{\gamma} \approx (B_{10}
I_\gamma)^{-1}$, where $B_{10}$ is the Einstein absorption
coefficient and $I_\gamma$ is the CMB intensity at the 21 cm
transition.  This yields $t_\gamma/t_H \sim 4 \times 10^{-6}
(1+z)^{1/2}$, where $t_H$ is the Hubble time.  If collisions or \lya absorption dominate, the timescale will
obviously be even smaller.}
\begin{equation}
T_S^{-1} = \frac{T_\gamma^{-1} + x_c T_K^{-1} + x_\alpha T_c^{-1}}{1 + x_{c} + x_\alpha}.
\label{eq:tsdefn}
\end{equation}
Here $x_c$ is the total collisional coupling coefficient, including
both H--H and H--e$^-$ collisions.  We will break the total coupling
into two components, $x_c^{\rm HH}$ and $x_c^{\rm eH}$, with obvious
meanings.  The coefficient from H--e$^-$ collisions is
\bq
x_c^{\rm eH} = \frac{n_e \kappa_{10} T_\star}{A_{10} T_\gamma},
\label{eq:xc}
\eq
where $n_e$ is the local electron density, $T_\star \equiv h
\nu_{21}/k_B = 0.068 \kel$, $\nu_{21}$ is the frequency of the 21 cm line, and
$A_{10}$ is the Einstein-$A$ coefficient for that transition.  The
last part of equation~(\ref{eq:tsdefn}) describes the
Wouthuysen-Field effect, in which the absorption and re-emission of
\lya photons mixes the hyperfine states.  The coupling coefficient
is \citep{chen04}
\begin{equation}
x_\alpha = 1.81 \times 10^{11} (1+z)^{-1} S_\alpha J_\alpha,
\label{eq:xalpha}
\end{equation}
where $S_\alpha \le 1$ describes the detailed atomic physics of the
scattering process and $J_\alpha$ is the background flux at the
Ly$\alpha$ frequency (ignoring scattering) in units of $\Junits$;
the Wouthuysen-Field effect becomes efficient when there is $\sim
0.1$ photon per baryon near this frequency.  It couples $T_S$ to an
effective color temperature $T_c$ (in most circumstances, $T_c
\approx T_K$; \citealt{field59-ts}).  Several estimates of
$S_\alpha$ and $T_c$ exist in the literature \citep{chen04,
hirata05, chuzhoy06, furl06-lyheat}.

\section{Electron-Hydrogen Collisions} \label{ehcoll}

\subsection{Mechanisms of collisional spin de-excitation} \label{mech}

In principle there are five distinct collisional mechanisms that can
cause spin de-excitation.  The existing literature has addressed
only one of these, electron spin exchange.  As we intend to increase
the accuracy to which $\kappa_{10}$ is known, we will first
reconsider the relative magnitudes of each mechanism.  We will then
calculate spin de-excitation cross sections and compare them to
previous results.

\subsubsection{Electron-electron spin exchange}
\label{collspin}

Electron-electron spin exchange refers to the process by which the
spin of the incoming electron is exchanged with that of the atomic
electron.  One such collision could be represented schematically by
\bq 
\alpha(s) + \beta(a)\alpha(p) \rightarrow \beta(s) + \alpha(a)\alpha(p), 
\label{eq:collspinex} 
\eq 
where $\alpha$ and $\beta$ refer to the Pauli spin states of spin-$\frac{1}{2}$
particles, $s$ and $a$ refer to the scattering and atomic electrons,
respectively, and $p$ refers to the proton.  In this collision, a
singlet hydrogen atom is converted into a triplet hydrogen atom by
spin exchange.  Including all possible spin permutations of this collision leads to
an expression for the spin de-excitation cross section.  (For a more
detailed discussion of spin exchange see \citealt{mott65} and
\citealt{condon63}.)

Ignoring all relativistic interactions and non-central forces, the
spin-exchange-mediated spin de-excitation cross section can be
calculated relatively simply.  (Interactions which violate these assumptions will be discussed
below in \S \ref{ee} and \S \ref{ep}.) We begin with the
time-independent Schr{\" o}dinger equation, which may be written 
\bq
\hat{H} \Psi({{\bf x}},{{\bf y}})=E \Psi({\bf x},{{\bf y}}),
\label{eq:schrod} 
\eq 
where, in atomic units, 
\bq
\hat{H}=-\nabla^2_{{\bf x}} - \nabla^2_{{\bf y}} - \frac{2}{|{\bf
x}|} - \frac{2}{|{\bf y}|} + \frac{2}{|{{\bf x}} - {{\bf y}}|},
\label{eq:ham} 
\eq 
$\Psi$ is the total wavefunction, and ${{\bf x}}$
and ${{\bf y}}$ label the positions of the two electrons. The Pauli
exclusion principle demands that 
\bq 
\Psi({{\bf x}},{{\bf y}}) = \pm \Psi({{\bf y}},{{\bf x}}), 
\label{eq:pauli} 
\eq 
where the spatially
symmetric and antisymmetric versions of the equation respectively
refer to singlet and triplet scattering.
We treat the scattering as electronically elastic (see \S
\ref{highn}), so the solution has the asymptotic form of 
\bq 
\Psi \sim e^{i\bk \cdot {{\bf z}}} + f_{\bf k}(\theta) \frac{e^{i{\bf k}
\cdot {\bf r}}}{r}, 
\label{eq:asymp} 
\eq 
where ${{\bf z}}$ is the direction of the incoming electron, ${{\bf k}}$ is its momentum, $r$
is the magnitude of the vector ${\bf r}$ separating the two
electrons, and $f_{{\bf k}}(\theta)$ is the scattering amplitude at
angle $\theta$.  Thus, the differential scattering cross section
into solid angle $\deriv \Omega$ is $\deriv \sigma/\deriv \Omega =
|f_{{\bf k}}(\theta)|^2$.  

We solve for the elastic scattering
amplitude by expanding the incoming electron into partial waves of
different orbital angular momentum (e.g., \citealt{mott65}),
yielding 
\bq 
f_{\bf k}^{s,t}(\theta)=\frac{1}{k} \sum_{L=0}^\infty
(2L+1) e^{i \delta_L^{s,t}} \sin \delta_L^{s,t} P_L(\cos \theta),
\label{eq:partial} 
\eq 
where $P_L$ is a Legendre polynomial of order
$L$, $\delta_L^{s,t}$ is the phase shift for angular momentum $L$ in
the singlet ($s$) and triplet ($t$) spin states, and $k$ is in units
of the Bohr radius $a_0$ (in this system energies are naturally
expressed in Rydbergs).  To obtain the total cross section, we
simply average over the initial spin states of the hydrogen atom, so
(in units of $a_0^2$) 
\bq 
\sigma_{\rm tot}({k}) = \frac{4 \pi}{k^2}\sum_{L=0}^\infty (2L+1) \left( \frac{1}{4} \sin^2
\delta_L^s + \frac{3}{4} \sin^2 \delta_L^t \right).
\label{eq:sigtot} 
\eq

The spin exchange cross section $\sigma_{\rm se}$ is more
complicated, because it is a coherent sum of the scattering
amplitudes \citep{field58, burke62b}:
\bq
\frac{\deriv \sigma_{\rm se}}{\deriv \Omega} = \frac{1}{16} | f_{{\bf k}}^t(\theta) -
f_{{\bf k}}^s(\theta) |^2,
\label{eq:diff-sigse}
\eq
so that (again in units of $a_0^2$)
\bq
\sigma_{\rm se}({k}) = \frac{\pi}{4 k^2} \sum_{L=0}^\infty (2L+1) \sin^2 ( \delta_L^t - \delta_L^s ).
\label{eq:sigse}
\eq

Thus, calculating the spin exchange cross section only requires
knowledge of the phase shifts for elastic scattering.  This is a
straightforward, though computationally challenging, problem.  As a
three-body interaction, no analytic solution exists, although one
can be found in the restricted problem of zero total and orbital
angular momentum \citep{temkin62, poet78}.  It has been solved
through variational principles \citep{schwartz61, shimamura71,
das76, register75, callaway78}, the close coupling formalism (in
which the total wavefunction is expanded in a basis constructed from
hydrogen eigenstates; \citealt{burke62a}), the related convergent
close coupling formalism \citep{bray02}, and through fully numeric
techniques \citep{wang93,wang94, shertzer94}.  All methods now agree
on $\delta_L^{s,t}$ to several significant figures in the range $0.1
\le k \le 0.8$; furthermore, these results agree with laboratory
experiments on spin-exchange effects in higher-energy H--e$^-$
collisions \citep{fletch85}.  We will use the results of
\citet{wang94}, who present the $L=0$--$3$ singlet and triplet phase
shifts.  Neither the $D$- nor the $F$-wave term contributes significantly to
$\sigma_{\rm se}$, so truncating the sum at $L=3$ is sufficient for
our purposes.

The cross section at zero energy requires more subtle analysis,
especially because the behavior at $k \le 0.1$ (or $E<0.01$ Ry) is
crucial for $T_K \la 1000 \kel$.  In the low-energy limit,
centrifugal barriers make the $L>0$ terms vanish, but the $S$-wave
scattering term remains finite.  The solution is typically presented
in terms of the scattering lengths $a_{s,t}$, defined so that $\tan
\delta_L^{s,t} \approx k a_{s,t}$.  Then, again in units of $a_0^2$,
\citep{seaton57} \bq \lim_{k \rightarrow 0} \sigma_{\rm se} =
\frac{\pi}{4} (a_t - a_s)^2. \label{eq:sigse-zero} \eq Scattering
lengths are difficult to compute because the effective potential
seen by the electron at zero energy dies off rather slowly.  The
most recent calculations appear to be by \citet{schwartz61}, who
found that $a_s=5.965 \pm 0.003$ and $a_t=1.7686 \pm 0.0002$.

We present detailed numerical results for the spin-exchange-mediated
spin de-excitation cross section in \S \ref{rate} below.  For comparison with other mechanisms, the magnitude of these cross sections in atomic units is of order unity.

\subsubsection{Electron-electron interactions}
\label{ee}

In addition to spin exchange, the incoming electron can interact
with the inherent magnetic dipole moment of the atomic electron.
(Note that spin-$\frac{1}{2}$ systems have only monopole and dipole
moments, so higher order multipoles need not be considered; \citealt{bethe36}.)

The incoming electron interacts with the magnetic moment of the
atomic electron by inducing a torque through a magnetic field.  The
interaction Hamiltonian, $\it {\hat H_{\rm int-e}}$, is given by \bq
{\it {\hat H_{\rm int-e}}}\ = - \frac{8 \pi \boldsymbol{\mu}_e^2
\delta(r)}{3} + \frac{1}{r^3} \left[ \boldsymbol{\mu}_e^2 - 3
\frac{({\bf r} \cdot \boldsymbol{\mu}_e)^2 } {r^2} - {\bf L} \cdot
\boldsymbol{\mu}_e \right], \label{eq:eehint} \eq where ${\bf L}$ is
the orbital angular momentum of the incoming electron,
$\boldsymbol{\mu}_e$ is the magnetic moment of the electron, given
by \bq \boldsymbol{\mu}_e  = \frac {g_e e \hbar} {4 m c},
\label{eq:emoment} \eq $g_e$ is the Land\'{e} factor for the
electron, the other symbols have their usual meanings, and we have
ignored any effect of the atomic nucleus \citep{jackson99}.  The
first three terms arise from the intrinsic magnetic moment of the
incoming electron, while the final term is generated by the motion
of the incoming electron and its associated charge.  A full solution
of this interaction would require relativistic quantum scattering
theory with non-central forces and is beyond the scope of this
study.

Fortunately, relativistic effects such as spin-spin and spin-orbit
coupling scale as $v^2 /c^2$ \citep{berestetskii82, goldman93}.  As
an upper limit, an electron with sufficient energy to ionize the
hydrogen atom (1 Ry) has velocity $v = 2 \times 10^8$ cm s$^{-1}$, so
$v^2 / c^2 = 5 \times 10^{-5}$.  (We will show below that this velocity corresponds to a temperature
above those at which our model is valid, so it suffices for a worst
case estimate.) Compared with the cross section of order unity due
to the spin-exchange mechanism, this is negligible.
The good agreement between the experimental results of
\citet{fletch85} and the cross sections calculated from
spin-exchange alone further supports our subsequent neglect of these
interactions.

\subsubsection{Electron-proton interactions}
\label{ep}

The incoming electron can also interact with the spin of the atomic
proton.  Again, this can occur in two different ways -- through the spin of
the electron and through the magnetic field generated by its motion.
The interaction Hamiltonian, $\it {\hat H_{\rm int-p}}$, is given by
\bqa {\it {\hat H_{\rm int-p}}} & = & - \frac{8 \pi
(\boldsymbol{\mu}_e \cdot \boldsymbol{\mu}_p)
\delta(r)}{3} + \nonumber \\
& & \frac{1}{r^3} \left[ (\boldsymbol{\mu}_e \cdot
\boldsymbol{\mu}_p) - 3 \frac{({\bf r} \cdot
\boldsymbol{\mu}_e)({\bf r} \cdot \boldsymbol{\mu}_p)} {r^2} - {\bf
L} \cdot \boldsymbol{\mu}_p \right], \label{eq:enhint} \eqa where
$\boldsymbol{\mu}_p$ is the magnetic moment of the proton, given by
\bq \boldsymbol{\mu}_p = \frac {g_p e \hbar} {4 m_p c},
\label{eq:nmoment} \eq $m_p$ is the mass of the proton, $g_p$ is the
Land\'{e} factor for the proton, and the other symbols are defined
as in equation~(\ref{eq:eehint}).

The same arguments used above for the electron-electron interaction
apply to the electron-proton interaction.  In this case, however, the magnetic moment of the nucleus is smaller than that of the electron by the ratio of their masses, $\approx
1/1836$, so the cross section is further suppressed to order
$10^{-8}$, and neglecting this interaction will not alter our
results.

\subsubsection{Transient complex formation}
\label{hminus}

In the region $k>0.8$, a series of resonances corresponding to
quasistable doubly excited states of H$^-$ occur.  These resonances
have two effects on our calculations.  First, the elastic scattering
phase shift undergoes an abrupt change by $\pi$ at a resonance. The
resonant phase shift can be fit to the usual Breit-Wigner form \bq
\tan (\delta - \delta_b) = \frac{\Gamma/2}{E_R-E}, \label{eq:res}
\eq where $\delta_b$ is the smoothly varying background phase shift,
$E_R$ is the center of the resonance, and $\Gamma$ is its width.  In
order to estimate the effects of this structure on the
spin-exchange-mediated spin de-excitation cross section, we include
the lowest order $^1S$, $^3P$, and $^1D$ resonances according to the
fits of \citet{wang94}.

Second, as the system passes through these resonances, it
transiently becomes H$^-$.  For a full treatment, we would need to
consider the evolution of the spin states during the time spent as
H$^-$ and the subsequent autodetachment reaction.  Two facts lead us
to ignore this possibility.  First, the resonances are broad.
The narrowest is the $^3P$ state, with a width of $2 \times 10^{-4}$
Ry \citep{wang94}; the corresponding natural lifetime of 12 fs is
much shorter than typical nuclear spin relaxation times. Second, the
resonances cover less than half of a percent of the energy range
considered. To the level of accuracy we seek, even complete spin
de-excitation at these resonances will not change our results.

\subsubsection{Electronic excitation of hydrogen}
\label{highn}

At the low temperatures relevant to the high-redshift IGM, nearly
all collisions occur below the $n=2$ excitation threshold.  However,
once this threshold is reached, excitation to higher-$n$ levels
quickly becomes energetically feasible.  Past this point, the plethora of possible transitions makes it
difficult to compute the H--e$^-$ collisional spin-exchange rates
from first principles.  We can, however, estimate the temperatures
at which such excitations can affect the de-excitation rate
significantly.  Detailed numerical results are given in \S \ref{rate}; the
conclusion is that a one percent effect is reached at $T_K \approx
1.5 \times 10^4 \kel$.

\begin{figure}
\begin{center}
\resizebox{8cm}{!}{\includegraphics{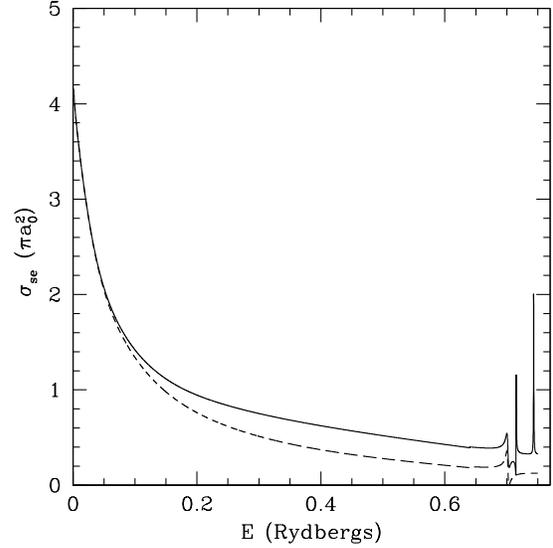}}\\%
\end{center}
\caption{Spin-exchange cross section (in units
of $\pi a_0^2$) for H--e$^-$ collisions, as a function of the energy
of the incident electron.  The solid curve shows $\sigma_{\rm se}$
including partial waves with $L \le 3$, while the dashed curve
includes only the $L=0$ term.  We include the three lowest energy
H$^-$ resonances.  } \label{fig:cs}
\end{figure}

\begin{figure*}
\begin{center}
\resizebox{8cm}{!}{\includegraphics{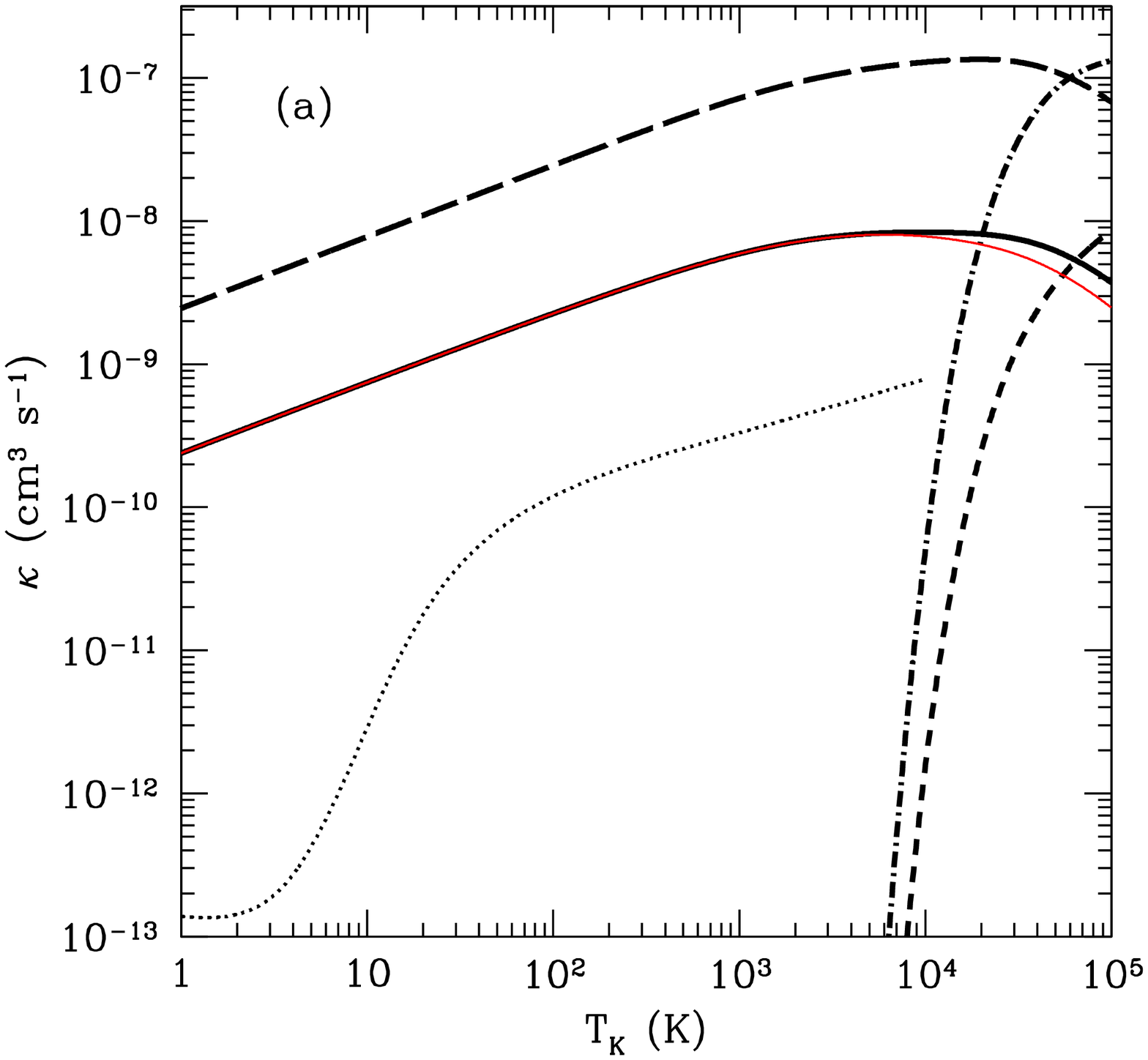}}
\hspace{0.13cm}
\resizebox{8cm}{!}{\includegraphics{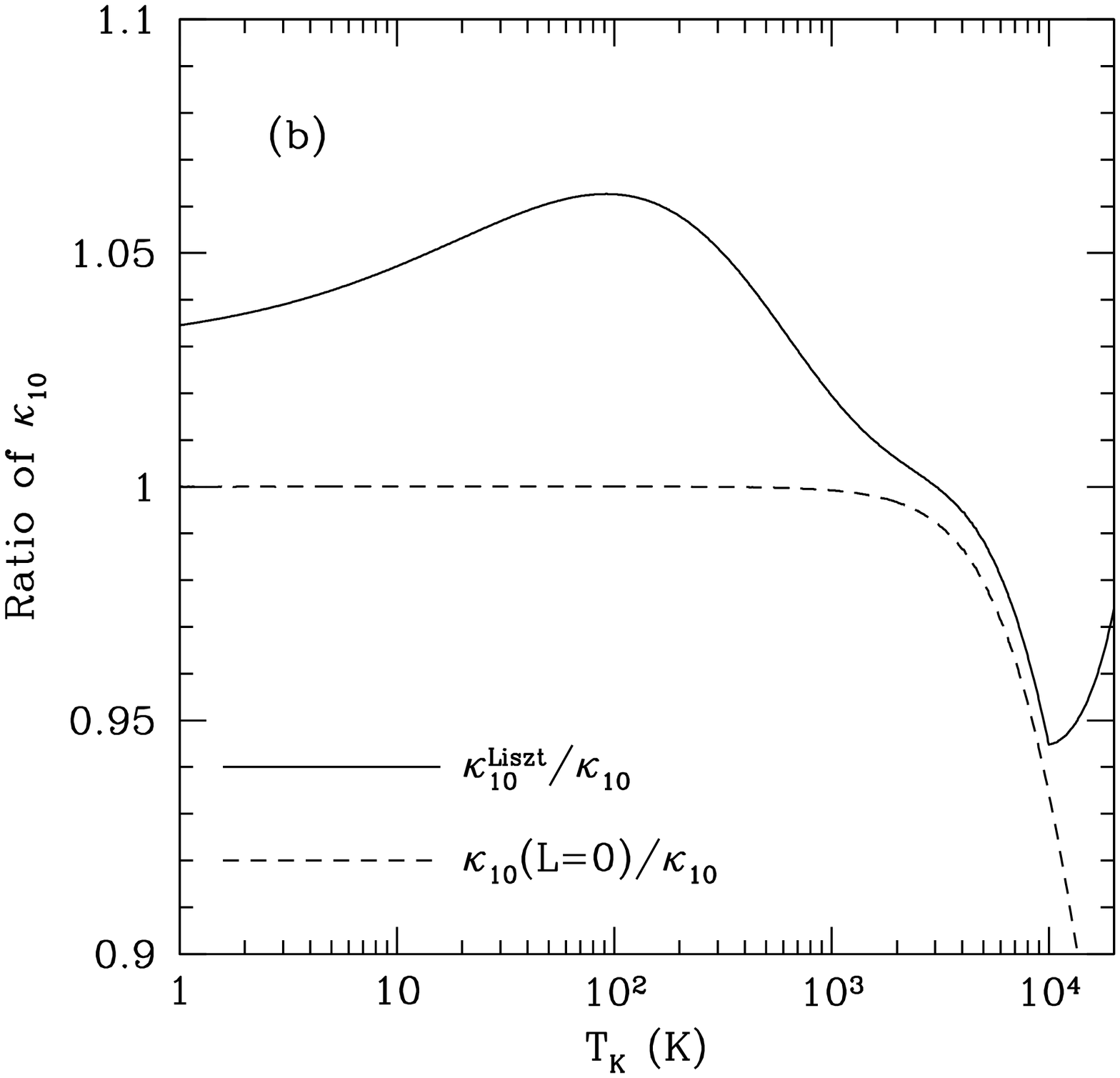}}
\end{center}
\caption{\emph{(a):}  Rate
coefficients.  The thick solid line shows $\kappa_{10}$, the spin
de-excitation rate from elastic collisions below the $n=2$
threshold.  The thin solid line shows the same quantity if we only
include the $L=0$ partial wave.  The long-dashed and dot-dashed
curves show $\kappa_{n<2}$ and $\kappa_{n \ge 2}$, respectively. The
short-dashed curve shows $\kappa_{2p}$.  For comparison, the dotted
curve shows $\kappa_{10}^{\rm HH}$.  \emph{(b):}  The solid curve
shows the ratio of the \citet{liszt01} fit for $\kappa_{10}$ to the
exact result.  The dashed curve shows the ratio of the $L=0$
solution to the full result.} \label{fig:kappa}
\end{figure*}

In practice, $\kappa_{10}$ quickly becomes irrelevant once the $n=2$
excitation threshold is reached, because the \lya background
generated by collisional excitations and subsequent radiative
de-excitations will then dominate the spin de-excitation process.
The production rate of \lya photons (in units of photons per volume
per second) is $\epsilon_\alpha = n_{\rm HI} n_e \kappa_{\alpha}$,
where $\kappa_\alpha$ is the rate coefficient for excitations that
eventually produce \lya photons and $n_{\rm HI}$ is the density of
hydrogen atoms.  For a simple estimate, we set $\kappa_\alpha =
\kappa_{2p}$, the rate of direct excitations to the $2p$ level.
This is a lower limit because of cascades from higher levels: note
that although roughly one-third of excitations to higher $p$ states
produce \lya photons \citep{hirata05, pritchard05, pritchard06}, the
cross sections for these excitations are significantly smaller than
the cross section for the $1s$--$2p$ transition, so the
approximation is justified. (For example, $\kappa_{3s}$ is
nearly fifty times smaller than $\kappa_{2p}$.)  Under this
approximation, the background flux at the \lya resonance (assuming
neutral helium) is \bq J_\alpha = \frac{c}{4\pi} \xhi (1 - \xhi)
\frac{n_H^2(z) \kappa_{2p}}{\nu_\alpha H(z)}, \label{eq:jalpha} \eq
where $\nu_\alpha$ is the frequency of the \lya transition.  In our
cosmology, the coupling coefficient is then \bqa
x_\alpha & \sim & S_\alpha \left( \frac{\kappa_{2p}}{2.5 \times 10^{-15} \recunits} \right) \left( \frac{1+z}{10} \right)^{7/2} \nonumber \\
& & \times \left[ \frac{\xhi(1-\xhi)}{0.25} \right].
\label{eq:xalpha-coll}
\eqa
Using the cross sections calculated below, we find that
$\kappa_{2p}$ is sufficiently large for \lya scattering to dominate
at $T_K \approx 6200 \kel$; see \S \ref{rate} for a more detailed discussion.

Overall, then, the effect of electronic excitations of hydrogen is
to limit the applicability of the present calculations at high
temperatures (above $T_K \approx 1.5 \times 10^4 \kel$).  Between
that limit and $T_K \approx 6200 \kel$, the effects of \lya photons
must also be included.

\subsection{The cross section} \label{cs}

Having considered all the possible mechanisms for spin de-excitation, we
have found that the spin-exchange mechanism does indeed dominate
under astrophysical conditions, at least to the fraction of a
percent level.  There are possibly some effects of H$^-$ resonances
in the region around $k = 0.8$ Ry, but otherwise our results are
valid to better than the percent level up to $T_K \approx 1.5 \times
10^4 \kel$.

In this section we proceed to calculate actual spin de-excitation
cross sections from the spin exchange rates.  Our inclusion of
higher-order partial waves is the principal improvement over
previous calculations.  \citet{field58} used the approximate
$S$-wave phase shifts of \citet{massey51}, which differ from the
true values by several percent over this energy range.  Most
significantly, their scattering lengths differ by $\sim 25\%$ from
the correct values.  \citet{smith66} used the (near-exact) $S$-wave
phase shifts of \citet{schwartz61} but did not include the $L>0$
terms.

Figure~\ref{fig:cs} shows $\sigma_{\rm se}$ as a function of
collision energy. The solid curve includes the $L=0$--$3$ terms,
while the dashed curve includes only $L=0$ and thus shows nearly the
same cross section used by \citet{smith66}.  Clearly the
higher-order partial waves become important at $k \ga 0.1$, where
non-zero angular momentum collisions become common.  In fact, the
$P$-wave makes up the vast majority of the difference (had we
included only $L=0$ and $1$, the solid curve would appear the same
except for the highest-energy resonance).  Note the resonance
structure at $k>0.8$; fortunately, although $\sigma_{\rm se}$
changes rapidly in this regime, the resonances constitute only a
small part ($\Gamma \la 3 \times 10^{-3}$ Ry) of the energy range.

\subsection{The rate coefficient} \label{rate}

The H--e$^-$ collisional spin de-excitation rate coefficient is \bq
\kappa_{10} = \sqrt{ \frac{8 k_B T_K}{\pi M}} \bar{\sigma}_{\rm se},
\label{eq:kappadefn} \eq where the prefactor is the mean collision
velocity, $M \approx m_e$ is the reduced mass of the H--e$^-$
system, and the thermally-averaged cross section is \bq
\bar{\sigma}_{\rm se} = \frac{1}{(k_B T_K)^2} \int_0^\infty \deriv E
\, \sigma_{\rm se}(E) E e^{-E/k_B T_K}. \label{eq:thermavg} \eq
Similarly, we can define $\kappa_{n<2}$, the rate coefficient for
\emph{all} collisions below the $n=2$ threshold, with the
replacement $\sigma_{\rm se} \rightarrow \sigma_{\rm tot}$.

The heavy solid line in Figure~\ref{fig:kappa}\emph{a} shows
$\kappa_{10}$ if we include only collisions below the $n=2$
threshold.  (In other words, we truncate the integral in eq.
\ref{eq:thermavg} at 10.2 eV; this restriction causes the decline at
$T_K \ga 40,000 \kel$.)  At low temperatures, it is nearly
proportional to $T_K^{1/2}$ (which comes from the velocity factor):
scattering is dominated by electrons near zero energy, where the
amplitude is essentially set by the scattering lengths $a_{s,t}$.
$\kappa_{10}$ flattens out at higher temperatures because
$\sigma_{\rm se}$ declines with energy (see Fig.~\ref{fig:cs}).  We
note that the resonances in Figure~\ref{fig:cs} are so narrow that
they never affect $\kappa_{10}$ by more than $1\%$.
In contrast, the dashed curve shows the total rate $\kappa_{n<2}$
for all collisions below the $n=2$ threshold.  Only $\sim 10\%$ of
collisions actually result in a net change of hydrogen spin.

For comparison, the dotted curve in Figure~\ref{fig:kappa}\emph{a}
shows the corresponding de-excitation rate coefficient for H-H
collisions.  The data at $T_K \la 300 \kel$ are taken from
\citet{zygelman05}.  At higher temperatures, we use cross sections
from K. Sigurdson (private communication, also tabulated in
\citealt{furl06-review}).  We do not show $\kappa_{10}^{\rm HH}$ at $T_K > 10^4 \kel$ because
Sigurdson did not include excitation to $n \ge 2$ states from H--H collisions.  At high temperatures, $\kappa_{10}/\kappa_{10}^{\rm HH} \sim
\sqrt{m_p/m_e}$ (comparable to the relative velocities of the
scattering species).  The hydrogen cross section is much smaller at
$T_K \la 100 \kel$ because of an accidental cancellation in the
$S$-wave cross sections \citep{zygelman05, sigurdson05-deut}.

The thin solid curve shows the cross section if we include only the
$L=0$ term (this is essentially identical to the result of
\citealt{smith66}); we also show the ratio between this version and
our result by the dashed curve in Figure~\ref{fig:kappa}\emph{b}.
As expected, higher-order partial waves only affect $\kappa_{10}$ at
$T_K \ga 1000 \kel$, where a substantial fraction of the electrons
have large enough momenta for the $L=1$ term to become significant.

Calculations in the literature often use the fit to $\kappa_{10}$
proposed by \citet{liszt01}, which is based on the high-temperature
results of \citet{smith66}.  The solid line in
Figure~\ref{fig:kappa}\emph{b} compares this function to the exact
value.  At low temperatures, it exceeds our results by several
percent, and it systematically underestimates $\kappa_{10}$ at $T
\ga 3000 \kel$.  We therefore recommend interpolating the exact rate
coefficients; to this end, Table~\ref{tab:kappa} presents our
results in numerical form.

\begin{table}
\begin{center}
\caption{Electron-hydrogen spin de-excitation rates}
\label{tab:kappa}
\begin{tabular}{|c|c|c|c|}
\hline
$T_K \, ({\rm K})$ & $\kappa_{10} \, (10^{-9} \, {\rm cm^3 \ s^{-1}})$ &$T_K \, ({\rm K})$ & $\kappa_{10} \, (10^{-9} \, {\rm cm^3 \ s^{-1}})$  \\
\hline
1 & 0.239 & 1000 & 5.92 \\
2 & 0.337 & 2000 & 7.15 \\
5 & 0.530 & 3000 & 7.71 \\
10 & 0.746 & 5000 &  8.17 \\
20 & 1.05 & 7000 & 8.32 \\
50 & 1.63 & 10,000 & 8.37 \\
100 & 2.26 & 15,000 & 8.29 \\
200 & 3.11 & 20,000 & 8.11 \\
500 & 4.59 & & \\
\hline
\end{tabular}
\end{center}
\end{table}

Unfortunately, there are no accurate calculations of $\sigma_{\rm
se}$ above the $n=2$ threshold; typically only the \emph{total}
cross sections for elastic scattering, excitation, and ionization
are presented (from which the spin-exchange cross section -- a
coherent sum of the singlet and triplet scattering amplitudes --
cannot readily be extracted).  However, we can at least estimate the
\emph{total} rate (including elastic scattering, excitation, and
ionization) for interactions above threshold (which we will call
$\kappa_{n \ge 2}$); comparing that to $\kappa_{n<2}$ provides an
estimate of the temperature at which our calculation breaks
down.  To compute $\sigma_{n \ge 2}$, we interpolate the
cross sections given by \citet{wang94} for energies between the
$n=2$ and $n=3$ thresholds (which include elastic collisions as well
as excitations to the $2s$ and $2p$ levels) and the total
cross sections for $E>13.6 \eV$ from the Convergent Close Coupling
online database\footnote{See
http://atom.murdoch.edu.au/CCC-WWW/index.html.} (see
\citealt{bray02} and references therein), which includes elastic
scattering, excitation to all levels through $4f$, and ionization.

The dot-dashed curve in the left panel of Figure~\ref{fig:kappa}
shows the resulting rate coefficient $\kappa_{n \ge 2}$.  It
increases rapidly at $T_K \ga 10^4 \kel$, when the tail of the
electron velocity distribution begins to populate the $E>10.2 \eV$
space.  We find $\kappa_{n \ge 2}/\kappa_{n < 2} \approx
0.01,\,0.1,$ and $1$ at $T_K \approx (1.5,\,2.3,\,6.0) \times 10^4
\kel$.  As mentioned above, we therefore expect that our
$\kappa_{10}$ is accurate to better than one percent at $T_K <1.5
\times 10^4 \kel$.

\begin{figure}
\begin{center}
\resizebox{8cm}{!}{\includegraphics{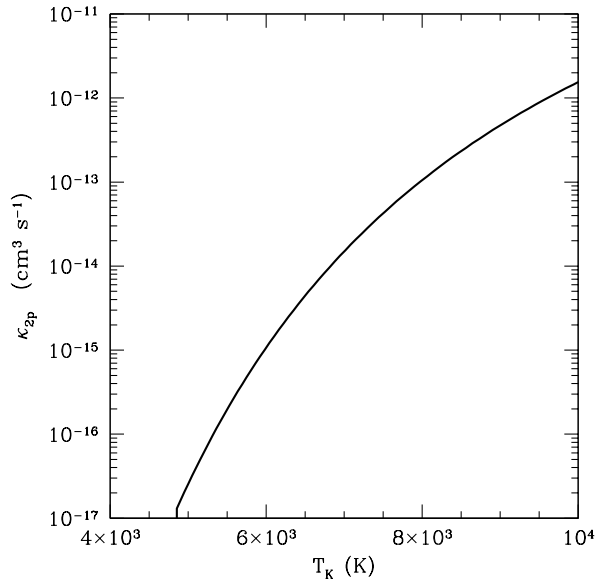}}\\%
\end{center}
\caption{Rate coefficient for H--e$^-$ collisional excitation to the $2p$ level.}
\label{fig:k2p}
\end{figure}

The short-dashed curve in Figure~\ref{fig:kappa}\emph{a} shows
$\kappa_{2p}$ (computed in the same way as the total cross section),
which we used in \S \ref{highn} above to calculate the temperature
at which \lya scattering becomes important; $\kappa_{2p}$ is
typically several percent of $\kappa_{n \ge 2}$.  Obviously it is an
extremely steep function of temperature, so we show a closeup of the
regime of interest in Figure~\ref{fig:k2p}.  We find that
$\kappa_{2p}$ reaches the level required by
equation~(\ref{eq:xalpha-coll}) at $T_K \approx 6200 \kel$.  At $T_K
\la 5000 \kel$, the \lya background can be neglected in most
applications; on the other hand, by $T_K \sim 8000 \kel$, it will
entirely dominate.  In detail, the total \lya production rate will
be slightly larger than shown here because of radiative cascades
from higher levels \citep{pritchard06}.  However, the extremely steep dependence of this rate coefficient on temperature suggests that in practice such minor corrections will
not be very important.  We also note again that Figure~\ref{fig:k2p}
assumes a thermal distribution of electrons; if a nonthermal
population of fast electrons exists (as is the case if X-rays
permeate the IGM; \citealt{chen06, chuzhoy06-first}), they can
collisionally excite higher levels and so produce \lya photons even
if the mean IGM temperature is much smaller.
Thus \lya production is probably always somewhat important, but its
details depend on the extremely uncertain radiation backgrounds
\citep{sethi05, furl06-glob}.  We will therefore not address this
possibility here.

\section{Astrophysical applications} \label{astro}

Although $\kappa_{10}$ is much larger than the corresponding
quantity for H--H collisions, H--e$^-$ collisions are often ignored
in calculating the spin temperature of the 21 cm line in the
high-redshift IGM.  At the small residual ionized fractions ($\bxion
\sim 2 \times 10^{-4}$) expected following standard cosmological
recombination \citep{seager99}, this is a reasonable assumption.
However, once $\bxion \ga 0.01$, electron collisions become quite
important (e.g., \citealt{nusser05-xray, kuhlen06-21cm}).  In this
section, we will apply our improved calculation of $\kappa_{10}$ to
some simple cosmological examples.  We will only consider the regime
$T_K \la 3000 \kel$, where the effects of electronic excitations of
hydrogen are much less than one percent and the \lya background from
collisional excitations can be ignored (assuming a Maxwell-Boltzmann
electron distribution and hence neglecting any X-ray background).
These examples are therefore not particularly realistic, but they
isolate the major effects of electron collisions and so are useful
model problems.

\begin{figure}
\begin{center}
\resizebox{8cm}{!}{\includegraphics{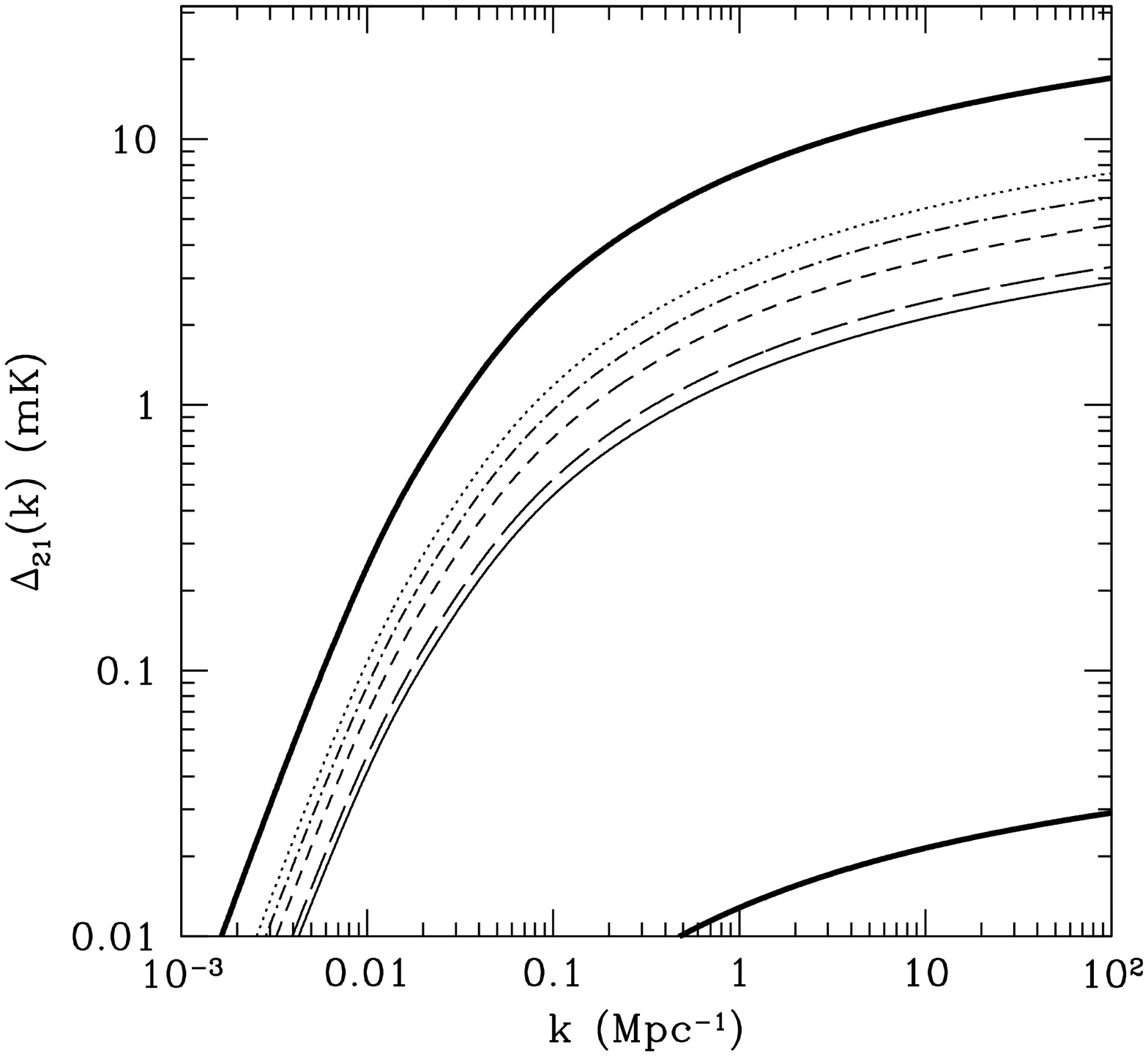}}\\%
\end{center}
\caption{21 cm brightness temperature fluctuations at $z=12$ for $T_K=10^3 \kel$.  The thin curves take $\bxion=0,\,0.01,\,0.05,\,0.1$, and $0.2$, from bottom to top, with $x_\alpha=0$.  The upper thick curve shows the fluctuation amplitude if $T_S=10^3 \kel$, while the lower thick curve shows the fluctuation amplitude if the temperature and ionization fraction are those expected in an adiabatically cooling Universe. }
\label{fig:pk}
\end{figure}

Equation (\ref{eq:dtb}) shows that perturbations in the density,
temperature, ionized fraction, and velocity all source fluctuations
in the brightness temperature.  Because, to linear order in
$k$-space, velocity perturbations are simply proportional to density
perturbations, we can write the Fourier transform
of the fractional 21 cm brightness temperature perturbation as \citep{furl06-review}
\begin{equation}
\tilde{\delta}_{21}(\bk) = (\beta + \mu^2) \tilde{\delta} + \beta_x \tilde{\delta}_x + \beta_T \tilde{\delta}_T,
\label{eq:d21}
\end{equation}
where $\tilde{\delta}$, $\tilde{\delta}_x$, and $\tilde{\delta}_T$
are the Fourier-space fractional perturbations in density, neutral
fraction, and $T_K$, respectively, and  $\mu$ is the cosine of the
angle between the line of sight and the wavevector $\bk$.  The
$\beta_i$ factors are linear expansion coefficients.  For
simplicity, we will assume that $\delta_T=0$ (see \citealt{pritchard06} for a detailed discussion of temperature fluctuations).  The relevant expansion coefficients $\beta_i$ are then
\citep{furl06-review}
\begin{eqnarray}
\beta & = & 1 + \frac{1}{1+x_{c}},
\label{eq:beta} \\
\beta_x & = & 1 + \frac{x_c^{\rm HH} - x_c^{\rm eH}}{x_{c} (1 + x_{c})}.
\label{eq:betax}
\end{eqnarray}

We assume that $\delta_x$ is determined by photoionization
equilibrium, \bq \delta_x = \left( \frac{1-\bxhi}{1+\bxhi} \right)
\delta \equiv g_x \delta_x, \label{eq:dx-pi} \eq where $\bxhi$ is
the global mean neutral fraction.  This is not a particularly good
assumption during reionization (again, see \citealt{pritchard06} for
a more careful treatment), but it allows us to compare our work with
previous results \citep{nusser05-xray}.  In any case, $\delta_x$
generally makes only a small contribution to the fluctuations.

From equation~(\ref{eq:d21}), it then follows that the
spherically-averaged power spectrum of 21 cm fluctuations can be
written
\bq
P_{21}(k) = \bdtb^2 (\beta'^2 + 2 \beta'/3 + 1/5) P_{\delta \delta}(k),
\label{eq:p21}
\eq
where $\bdtb$ is the mean brightness temperature, $P_{\delta \delta}(k)$ is the matter power
spectrum, and $\beta' = \beta + \beta_x g_x$.  We will quote our
results in terms of the mean temperature fluctuation as a function of scale,
$\Delta_{21}^2(k) = (k^3/2\pi^2) P_{21}(k)$.

\begin{figure}
\begin{center}
\resizebox{8cm}{!}{\includegraphics{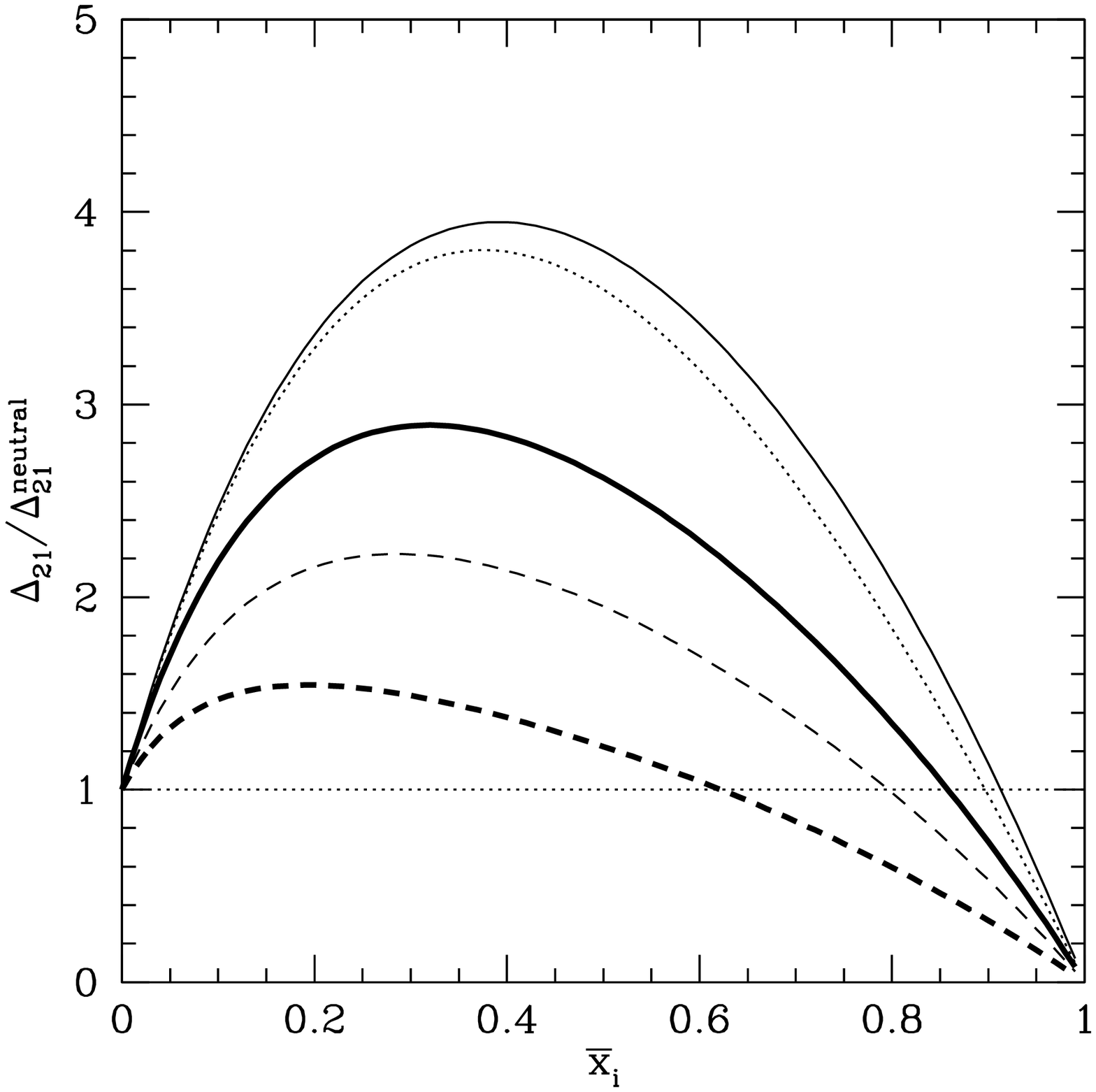}}\\%
\end{center}
\caption{Ratio of the 21 cm brightness temperature fluctuation amplitude to its value in a purely neutral Universe, as a function of $\bxion$.  The thin and thick curves are for $z=12$ and $z=20$, respectively. The solid and dashed lines take $T_K=100$ and $3000 \kel$.  The dotted curve shows the fluctuations at $z=12$ and $T_K=3000 \kel$ when $\delta_x=0$.  }
\label{fig:pkratio}
\end{figure}

Figure~\ref{fig:pk} shows the brightness temperature fluctuations in
some example scenarios at $z=12$.\footnote{Note that here we use the
linear version of the matter power spectrum, $P_{\delta \delta}$,
with the transfer function of \citet{eisenstein98}.  At the smallest
scales in the plot, nonlinear corrections will actually be
significant (e.g., \citealt{iliev03, furl04-bub, naoz05}).}  The
lower thick solid curve shows $\Delta_{21}$ for the standard
calculation, in which the only heat source is Compton scattering of
CMB photons (computed with RECFAST; \citealt{seager99}).  In this
case the gas is cold ($T_K=3.51 \kel$) and almost entirely neutral
($\bxion = 1.8 \times 10^{-4}$), so both collisional coupling and
the 21 cm fluctuations are extremely weak.  The upper thick solid
curve shows a contrasting case in which $T_S=10^3 \kel$ and
$\bxion=0$.  Here the IGM has saturated in emission.

The thin curves assume $T_K=10^3 \kel$ and take
$\bxion=0,\,0.01,\,0.05,\,0.1$, and $0.2$, from bottom to top.
Naively, of course, one would expect $\Delta_{21}$ to decrease as
$\bxion$ increases, simply because there is less neutral gas.  But,
as we have seen, H--e$^-$ collisions are much  more efficient at
changing $T_S$ than H--H interactions, so they can enhance the
fluctuation amplitude by more than a factor of two even at these
relatively small amounts of ionization.  As a result, warm gas can
seed milliKelvin fluctuations even in the absence of a \lya
background (regardless of the heat source).

Figure~\ref{fig:pkratio} shows how the amplification factor depends
on $\bxion$ by comparing to the fluctuation amplitude in fully
neutral gas.  First consider the two thick curves, which show the
ratio between the signals with and without ionization at $z=20$. The
solid and dashed curves assume $T_K=100$ and $3000 \kel$,
respectively.  The amplification is larger for cooler gas because
the overall coupling is weaker at lower temperatures, and the
addition of free electrons makes more of a difference.
Interestingly, electron coupling provides a large enough boost that
ionizing the gas continues to increase $\Delta_{21}$ until $\bxion$
is quite large -- even as high as $90\%$ at the lower temperature.

The thin curves show the same ratios at $z=12$.  Here the
amplification is even larger because the densities are smaller (and
hence the coupling weaker for a given $T_K$ and $\bxion$).  For
$T_K=100 \kel$, we also show the ratio assuming $\delta_x=0$
(rather than determined by photoionization equilibrium) with the
dotted curve.  Obviously, variations in the ionized fraction play
only a minor role compared to the density fluctuations.

Our results can be compared directly with those of
\citet{nusser05-xray}, who presented more detailed calculations of
21 cm fluctuations in a warm Universe (in his case, one flooded with
X-ray photons).  He used the $\kappa_{10}$ values calculated by
\citet{field58}, following the approximate phase shifts of
\citet{massey51}.  These overestimated the scattering lengths by
$\sim 25\%$, so our predicted amplitudes are somewhat smaller than
his.  Note again that, because we have neglected fast photoelectrons
from X-rays \citep{chen06, chuzhoy06-first}, as well as temperature
fluctuations \citep{pritchard06}, these Figures are probably too
simplistic to offer anything more than basic intuition.

\section{Discussion} \label{disc}

We have computed the spin de-excitation rates for hydrogen in the
ground state during electron-hydrogen collisions.  Previous
calculations assumed that spin exchange through elastic $S$-wave
scattering was the only relevant mechanism.  Our analysis showed
that while spin exchange does dominate spin de-excitation, higher
partial waves contribute significantly to the total de-excitation
rate.  We used newer calculations of the elastic scattering phase
shifts for the $L=0$--$3$ partial waves \citep{wang94} to re-compute
$\kappa_{10}$.  Our main results are presented in
Table~\ref{tab:kappa}.  They remain accurate up to $T_K \la 1.5
\times 10^4 \kel$, where collisions above the $n=2$ threshold begin
to become important at the percent level.  Our results differ from
\citet{smith66} and especially from the widely-used fit in
\citet{liszt01} by several percent over the range $1 \kel < T_K <
10^4 \kel$.

We have also shown that spin coupling produced directly through
H--e$^-$ collisions will become a secondary effect at $T_K \ga 6200
\kel$, because \lya photons produced through collisional excitations
easily dominate the spin coupling in this regime.  Although the
excitation rate to the $2p$ level (and to other levels that cascade
through Ly$\alpha$) can be much smaller than the collisional spin
de-excitation rate, each \lya photon scatters $\sim 10^5$ times
before redshifting out of resonance.  This dramatically boosts the
efficiency of the radiation background relative to collisions.  In
practice, the \lya background may be important at even lower
temperatures.  In our calculations, we have assumed a
Maxwell-Boltzmann distribution of electron velocities, but even a
small population of nonthermal high-energy electrons can induce
significant \lya coupling.  For example, X-rays produce fast
secondary electrons able to excite hydrogen collisionally, which can
easily produce such a background \citep{chen06, chuzhoy06-first}.

We also considered some simple applications of our results to the
high-redshift IGM.  Electron-hydrogen collisions are likely to be
most important if the radiation background is dominated by X-rays,
so that the IGM becomes warm and weakly ionized.  Such scenarios are
particularly common when black hole accretion provides a significant
fraction of the ionizing flux \citep{ricotti04_a, ricotti05,
kuhlen06-21cm}.  We have seen that  the typical 21 cm fluctuation
amplitude can reach the milliKelvin level even without a significant
ultraviolet background.  Our rate coefficients will be useful in
assessing the implications of such scenarios for the 21 cm sky.

\vspace{0.1cm}

SRF thanks the Tapir group at Caltech for their hospitality while
much of this work was completed.  This publication has been approved for release as LA-UR-06-5395.  Los Alamos National Laboratory, an affirmative action/equal opportunity
employer, is operated by Los Alamos National Security, LLC, for the
National Nuclear Security Administration of the U.S. under contract
DE-AC52-06NA25396.


\end{document}